\documentclass[a4paper,preprint]{revtex4}
\usepackage{hyperref}

\begin{document}


\title{Towards an anisotropic atom-atom model for the crystalline phases of the
molecular S$_8$ compound}

\author{C. Pastorino and Z. Gamba}
\affiliation{Department of Physics, Comisi\'on Nacional de Energ\'\i a 
At\'omica-CAC,\\
 Av. Libertador 8250, (1429) Buenos Aires, Argentina}
\email{clopasto@cnea.gov.ar, gamba@cnea.gov.ar}

\begin{abstract}
 
We analize two anisotropic atom-atom models used to describe 
the crystalline $\alpha $, $\beta $ and $\gamma $ phases of 
S$_8$ crystals, the most stable compound of elemental sulfur in 
solid phases, at ambient pressure and T$_{\sim}^<$400 K. The 
calculations are performed $ via $ a series of classical molecular 
dynamics (MD) simulations, with flexible molecular models and 
using a constant pressure-constant temperature algorithm for the 
numerical simulations. All intramolecular modes that mix with 
lattice modes, and are therefore relevant on the onset of 
structural phase transitions, are taken into account. Comparisons 
with experimental data and previous results obtained with an 
isotropic atom-atom molecular model are also performed.\\

\end{abstract}
                              
\maketitle
\section{Introduction}
 
Our knowledge of the complex phase diagram of elemental sulfur, 
as well as of the statical, dynamical and thermodynamical 
properties of all its condensed phases, is far from 
complete \cite{steudel,meyer1}. Its complexity is due to a 
rich variety of molecular allotropes (open and cyclic chains of 
different lengths) that correlate with a large variety of
chemical and physical properties in the crystalline and liquid 
phases \cite{steudel,meyer1}.\\

 The most stable, and the most studied, sulfur allotrope at ambient 
temperature and pressure (STP) is S$_8$, a crown-shaped cyclic 
molecule.  Three crystalline phases, $\alpha $-, $\beta $- and 
$\gamma $-S$_8$, have been reported at ambient pressure and 
 T$_{\sim}^<$400 K. Figs. 1a, b and c show their structures.
The  orthorhombic $\alpha $-S$_8$ can be measured \cite{alfa1,alfa2} 
at temperatures up to 385.8 K, its metastable melting point. 
 At 369 K there is a solid-solid  $\alpha $-$\beta $ phase 
transition, but it is most often observed when the samples are 
studied from high to low temperatures.
 Monoclinic $\beta $-S$_8$ \cite{beta2} is the most stable 
phase in the 369-393 K temperature range. $\beta $-S$_8$ crystals 
are usually obtained from the melt \cite{steudel} and they 
show an orientational order-disorder phase transition 
at 198 K \cite{beta1}.
 A third crystalline allotrope $\gamma $-S$_8$ \cite{gam1} 
(Fig. 1c) has been observed, with a density, at STP, 5.8 \% higher 
than that of $\alpha $-S$_8$.\\

 Most theoretical studies of the S$_8$ crystalline phases, 
up to very recently, have been limited to calculations of the 
intramolecular mode frequencies for the isolated S$_8$ molecule 
and their splitting when they are embedded in a $\alpha $-S$_8$ 
matrix, including calculations of their infrared and Raman 
intensities \cite{alfafreq}. Constant-volume molecular 
dynamics (MD) simulations with the same aim have been also 
performed \cite{cardini1}.\\

 We recently reported \cite{claudio1,claudio2} a series of 
constant pressure-constant temperature classical MD simulations 
of $\alpha-,$ $\beta-$ and $\gamma-S_8$ crystals using an 
extremely simple and flexible molecular model, a slight 
modification of the one proposed in the constant volume MD 
simulations of $\alpha $-S$_8$ \cite{cardini1}. The simple 
intermolecular potential was of the Lennard-Jones (LJ) atom-atom 
type, except that its parameters were fitted in 
ref. \onlinecite{claudio1} to the experimental data on the STP 
crystalline structure of $\alpha $-S$_8$ and an estimated value 
of its configurational energy, calculated as in 
refs. \onlinecite{claudio1,claudio2} and based on the experimental STP 
value of the heat of sublimation of $\alpha $-S$_8$ \cite{hsubalfa}
(it is also explained in the Calculations section of this paper). 
These LJ parameters 
($\varepsilon $$_{iso}$= 1.70 kJ/mol and $\sigma $$_{iso}$=3.39 \AA) 
were afterwards used throughout the calculations of all samples 
of $\alpha-,$ $\beta-$ and $\gamma-S_8$ crystals at different 
temperatures and zero kbar pressure \cite{claudio1,claudio2}.
The intramolecular potential included a harmonic well term for the 
SSS angle and a torsional potential of the double well type for the 
sulfur chain, different from hydrocarbon chains that show a 
triple well torsional potential. These intramolecular coordinates
fully describe all the low frequency intramolecular modes that mix 
with lattice modes \cite{alfafreq,cardini1} and can, therefore, be 
relevant for the lattice stability or in the onset of structural 
phase transitions.\\ 

 Our calculations showed that the simple and flexible molecule 
model succesfully reproduced most experimental data available on 
the three crystals \cite{claudio1,claudio2}, including structural 
and dynamical data of $\alpha $- and $\gamma $-S$_8$ at STP, the 
high temperature orientationally disordered phase of $\beta $-S$_8$ 
at 375K and its low temperature ordered phase.
Nevertheless, below 200 K this simple isotropic LJ model did not
reproduce the 100 K structure of $\alpha $-S$_8$, as determined 
by a neutron and X-ray measurements \cite{alfa3}. Although no 
structural phase transition was measured for $\alpha $-S$_8$ 
crystals \cite{alfa3}, when decreasing the temperature of the 
sample below 200 K, our orthorhombic sample turned out 
to be unstable using this isotropic atom-atom potential, 
and transformed to a monoclinic structure that we labeled
 $\alpha $'-S$_8$ \cite{claudio1,claudio2} (Fig. 1).\\

   Assuming that the experimental data of ref. \onlinecite{alfa3} do not 
correspond to a metastable state, the problem can be correlated to 
the extremely simple model molecule and in this work we propose to 
take into account the anisotropy of the sulfur atom due to its 
3s$^2$ 3p$^4$ electronic configuration. This idea is not new, it 
gave excellent results in similar problems. For example, in the case 
of alkane hydrocarbon chains it is known that an unique isotropic 
atom representing the CH$_2$ units is able to reproduce many facts 
of the high temperature phases \cite{alkane2}, but agreement with 
the measured crystalline structures is attained only when hydrogen 
atoms are explicitly included \cite{alkane1}. An intermediate
model was proposed in ref. \onlinecite{alkane3}, consisting in displacing 
an unique isotropic interaction site from the C position to the center 
of mass of the CH$_2$ unit, this improvement was enough to study the 
liquid phase of alkanes but not their crystalline phases \cite{alkane4}. 
The "all atoms" model has afterwards been succesfully applied to 
many studies, like the conformations of flexible 
molecules \cite{chains}, the equation of state of alkanes
\cite{alkane4} and the condensed phases of glycerol \cite{glycerol}.\\

Also in ref. \onlinecite{williams} the parameters of an aspherical atom-atom 
model were fitted to reproduce 11 crystalline structures of elemental 
sulfur, the location of the sites that simulate the lone-pair 
electronic distribution was refined. The calculations were performed 
in the harmonic approximation, for rigid molecules, and no attempt 
to include temperature, intramolecular soft modes, lattice stability 
or vibrational frequencies was performed.\\

   In the following sections we present the models of the anisotropic 
intermolecular potential that are analized in the present calculations,
the method of calculation, the obtained results for $\alpha$-, 
$\alpha$'-, $\beta-$ and $\gamma-S_8$ crystals and our conclusions.\\

\section{The anisotropic atom-atom potential}  

  For the present calculations we analize two simple anisotropic
 atom-atom LJ intermolecular potential models, based on the 
experimental charge distribution of $\alpha $-S$_8$ \cite{alfa3}, 
where the location of the sulfur lone pair orbitals were 
approximately determined. 
Our first anisotropic case molecule model 
model of refs. \onlinecite{claudio1,claudio2} replaces each isotropic 
LJ atomic site by four sites at a distance $d_{ani}$ from the S atom, 
two of them located along the SS bonds and the other two are
generated with the S$_4$ element of symmetry located at the atomic 
S site. This multi-site model (that we called ANI-S$_4$) is a simple 
way to represent the atomic anisotropy. \\
For the distance $d_{ani}$ we considered two values. The first one is 
$d_{ani}$=0.6 \AA, the experimental value determined at 100K \cite{alfa3}.
The second is a reduced value $d_{ani}$=0.1 \AA, that was 
included after taking into account that our four sites anisotropic 
model is extremelly simple and do not include a site at the S atom, 
making artificially large the anisotropy of the S atom.

The new LJ parameters of these sites were adjusted to fit the 
experimental structure and the estimated value of the potential 
energy of $\alpha $-S$_8$ at 300K, obtaining   
($\varepsilon $$_{ani}$=0.225 kJ/mol, $\sigma $$_{ani}$=2.87 \AA)
for $d_{ani}$=0.6 \AA, and 
($\varepsilon $$_{ani}$=0.115kJ/mol, $\sigma $$_{ani}$=3.35 \AA) 
for $d_{ani}$=0.1 \AA.
It has to be taken into account that, should we have taken
$d_{ani}$=0\AA, the LJ parameter's values should have been: 
$\varepsilon $$_{ani}$= $\varepsilon $$_{iso}$/16=0.107kJ/mol and
$\sigma $$_{ani}$=$\sigma $$_{iso}$=3.39 \AA.\\

As we show in the following sections, due to the results obtained 
with the ANI-S$_4$ potential model, for both $d_{ani}$ distances, 
a further an extremely anisotropic atom model was tested, for the 
sake of completeness. The last model locates the two external 
sites almost perpendicular to the plane determined by the 
corresponding S atom and the two nearest S neighbors, at
$d_{ani}$=0.6 \AA\ from the S atom (ANI$_\perp $ model). 
This approximated location is also suggested by the experimental 
data of ref. \onlinecite{alfa3} and the MO calculations that they 
include, on the sulfur containing simple molecule H$_2$S$_2$. 
As we explain in the next section, the most stable numerical 
simulations were obtained when the two sites are generated
by simplely displacing the two sites at a distance 
$d_{ani}$=0.6 \AA\ 
from the location of the corresponding atom {\bf S}$_j$ (j=1,8) 
and orienting them along the vectors $^+_-$[({\bf S}$_{j+2}$
-{\bf S}$_{j+1}$)+({\bf S}$_{j-2}$-{\bf S}$_{j-1}$)].
That is, the two external sites are located in a plane parallel 
to that formed by the next two nearest SS bonds. This direction 
is approximately perpendicular to the plane formed by the 
S$_{j-1}$S$_j$S$_{j+1}$ atoms and this model is considered a 
case study that represents an extremely large atomic anisotropy. 
The LJ parameters of the ANI$_\perp $ model were also adjusted
to fit the experimental structure and estimated potential
energy of $\alpha $-S$_8$ at STP, obtaining in this case  
$\varepsilon $$_{ani}$=0.15 kJ/mol and 
$\sigma $$_{ani}$=3.17 \AA.\\

The LJ parameters of each multisite model, all of them adjusted
to fit the experimental STP data of $\alpha $-S$_8$, were 
afterwards used to perform the series of MD runs of the above 
mentioned four crystalline phases, at zero pressure and in the
50K - 450K temperature range. \\

 The intramolecular terms of the potential model, for all molecular 
models, are essentially equal to those of refs. 
\cite{claudio1,claudio2}, with the only difference that the force 
constants take now a value about 10 \%\ lower, in order to improve 
the fit of the intramolecular frequencies to the experimental data. 
The intramolecular potential for the S-S-S bending angles consists 
of a harmonic well

\[
V(\beta )=\frac 12C_\beta (\beta -\beta _0)^2, 
\]

with a force constant of C$_\beta $/k$_B$=22000 K/rad$^2$ and $\beta _0$= 
108 deg., k$_B$ is the Boltzmann constant. The intramolecular 
potential for torsion ($\tau $) angles is a double well:

\[
V(\tau )=A_\tau +B_\tau \cos (\tau )+C_\tau \cos ^2(\tau )+D_\tau \cos
^3(\tau )\text{,} 
\]

with $A_\tau $/k$_B$=51.473 K, $B_\tau $/k$_B$=664.573 K,
 $C_\tau $/k$_B$=2068.092 K and $D_\tau $/k$_B$=476.453 K. These 
parameters describe a double well with minima at 
$\tau $=${^+_-}$ 98.8 deg., and a barrier height of about 8 kJ/mol at 
$\tau $=180 deg. The barrier height at $\tau $=0 deg. is of 27 kJ/mol, out of
the range of energies explored in these simulations.\\
The bond lengths S-S are held constant at their mean experimental value 
of 2.06 \AA.

\section{Calculations} 

The method of calculation, as well as the performed series of numerical
simulations are equal to those included in refs. \onlinecite{claudio1,claudio2},
the only change is in the potential model used.\\

 Briefly, the phase diagram and dynamical bulk properties of S$_8$ crystals, 
as given by these simple molecular models, are calculated at 0 kbar 
pressure in the (N,P,T) ensemble, by a series of classical MD simulations. 
 The external pressure is considered isotropic and the MD algorithm allows
volume and shape fluctuations of the sample, in order to balance the
internal stresses with an external isotropic pressure \cite{algor2}. The 
algorithm controls pressure $ via $ an extended system which includes as
extra variables the MD box parameters, periodic boundary conditions are applied. The 
temperature control of the 
sample follows the approach of Nos\'e \cite{algor4,algor5}, which also 
includes an external variable. The equations of motion of these 
flexible molecules are integrated using the Verlet algorithm for the 
atomic displacements and the Shake algorithm for the constant bond length  
constraints on each molecule \cite{algor1}. The MD\ algorithm 
is based in  that used in a study of black Newton 
films \cite{bubbles1}, afterwards extended to constant-pressure constant-temperature 
simulations via an extended Lagrangian \cite{newsimul1,newsimul2}. The final version is 
similar to that used in ref. \onlinecite{zully-c4f8}.  
In the present work, as the 
interaction sites of our anisotropic atom model are not coincident 
with the S atomic sites, the algorithm employed to translate 
the forces from massless to massive sites is based on that written 
for rigid molecules\cite{algor6}, extended to the case of flexible molecules.

The runs in the (N,P,T) ensemble were performed every 25 or 50 K in the
range  50-450 K. At each point of the phase diagram the samples are 
equilibrated for 20000 to 30000 time steps (each one of 0.01ps.) and 
measured in the following 20 ps. Near the phase transitions the 
stabilization times were increased several times.\\
The exception are the series of runs corresponding to the ANI-S$_4$
model with $d_{ani}$=0.6\AA, that were performed with 
 a time step of 0.005ps. Nevertheless, these runs, after equilibration 
of the sample at the desired T and P, and in a free trajectory in 
the phase space, give a large deviation of the total energy
of the system (including the aditional coordinates of the extended
system), implying that the numerical algorithm should be improved
or that the molecular model do not calculate the sample as a
stable one. For this model we only include, for the sake
of comparison, the calculated structure 
and configurational energy that is measured (after the initial
transient) in the equilibration
trajectory, an algorithm that following ref. \onlinecite{algor3} 
simulates the system in the canonical ensemble.\\

The samples of $\alpha $-S$_8$ and $\alpha $'-S$_8$ crystals 
consisted of 3*3*2 orthorhombic cells (288 molecules).
The sample of $\beta $-S$_8$ crystals consisted of 4*4*4 cells (384 
molecules) and that of $\gamma $-S$_8$ crystals of 5*3*5 cells (300
molecules).\\ 

The free energy of all samples is estimated, at each point of the 
phase diagram, taking into account the contribution of the calculated 
vibrational modes to the entropy of the sample \cite{bornhuang}. 
This simple method is widely used in lattice dynamics calculations: 
In a first step the calculated vibrational density of states VDS($\nu $)
is normalized to the total number of degrees of freedom of the MD 
sample.  This normalized function VDS($\nu $) gives the number of 
oscillators within $\nu $$^{+}_{-}$$\Delta \nu $ and in a first 
approximation, by considering a system of harmonic oscillators, 
the calculation of their contribution to the entropy is 
straightforward \cite{bornhuang}. In the same way the internal 
energy and enthalpy of the system can be derived, the enthalpy
being directly comparable to thermodynamic measurements.

\section{Results}  

Using the anisotropic LJ model ANI-S$_4$, with distances
$d_{ani}$=0.1 and 0.6\AA, the orthorhombic $\alpha $-S$_8$ 
sample does not spontaneously show the structural phase transition 
to the monoclinic $\alpha $'-S$_8$, as was calculated with the 
isotropic LJ model when the temperature of the orthorhombic 
sample was decreased below 200 K \cite{claudio1,claudio2}. 
Nevertheless, if a new series of simulations are performed 
on a sample that initially has the structure of the 
monoclinic $\alpha $'-S$_8$, this phase is calculated stable, in the time scale of
our simulations (about 200 ps.), with a potential energy lower than that of $\alpha $-S$_8$. 
Fig. 2 shows the calculated configurational energy (the time 
and sample average of the potential interaction energy per 
molecule) and volume per molecule as a function of temperature, 
for the four crystalline samples, and using the ANI-S$_4$ model 
with $d_{ani}$=0.1 \AA. Table I includes our results for the 
ANI-S$_4$ model with $d_{ani}$=0.1 and 0.6 \AA\ at selected 
points of the phase diagram. \\ 

 A dramatic change is observed when the ANI$_{\perp}$ potential
model is used. These results are included in Fig. 3 and they 
clearly show a large diference with those of Fig. 2. The extremely 
anisotropic case model ANI$_{\perp}$ gives results that, in general,
 compare better with some of the experimental data than those included in Fig. 2.
 In particular, Fig. 3 shows that 
 the low temperature orthorhombic $\alpha $-S$_8$ phase
 is now calculated with a lower configurational energy
(and volume per molecule) than the monoclinic
 $\alpha $'-S$_8$ phase suggested by the isotropic and ANI-S$_4$ 
models. Nevertheless, this phase is calculated
as stable with all tested models and this fact implies that 
$\alpha $'-S$_8$ might be a metastable phase of elemental sulfur.\\
As we have also discussed in ref. \onlinecite{claudio1}, the uncertainty
in our comparison with the scarce experimental data on
$\beta $- and $\gamma $-S$_8$ phases
is mainly due to contradictory measurements of 
$\gamma $-S$_8$ crystals. Density measurements at 300K imply that
their density is 5.8\%\ higher than that of $\alpha $-S$_8$ at 
STP \cite{meyer1,gam1}. On the other hand the X-ray 
measurements \cite{gam1} determine a crystalline cell with a
larger volume per molecule than that of $\alpha $-S$_8$,
implying a 2.3\%\ lower density. Further measurements of
this phase are needed.\\

 The orientationally disordered phase of
$\beta $-S$_8$ is reproduced by the isotropic \cite{claudio1},
the ANI-S$_4$ and ANI$_\perp $ anisotropic models.
 At high temperatures, the molecules at (0,0,0) and 
(1/2,1/2,1/2) are dynamically disordered. The inertial axis
 perpendicular to the molecular plane librates around its equilibrium
orientation, whilst the molecule reorientates within
the plane. We calculate a characteristic reorientational time (within
molecular plane) of 9 ps at 375 K with ANI-S$_4$ ($d_{ani}$=0.6\AA)
 and of about 50 ps at 375 K with ANI$_\perp $. Although no
sharp transition is found for this crystal when decreasing the
temperature of the sample, below 150 K both 
molecules are orientationally ordered and related by a C2 symmetry 
axes. No measurements on the molecular reorientational times and
lattices parameters at diferent temperatures are available
for comparison. \\

  It has to be pointed out that, when analyzing the relative stability
of two phases we have estimated the Gibbs free energy of our samples
by including the contribution of the vibrational modes to the
entropy. Fig. 4 compares the data of $\alpha $-, $\alpha $'-,
$\beta $-  and $\gamma $-S$_8$. At low temperatures the relative
stability of the phases is equal to that obtained from the 
potential energy data. At high temperatures the merge of all phases 
is observed. No crossings (which identify phase transitions) are
found.\\
                                    
Table I compares the experimental and calculated (using the isotropic 
and anisotropic models) structural data at the starting point of 
the phase diagram of our series of
simulations. The calculated average pressure, not included in the Table is
 0 kbar in all cases. For example, for $\alpha $-S$_8$ at 300K
is P=0.0(5) kbar.\\
                          
   As in refs. \onlinecite{claudio1,claudio2}, the $\alpha $- $\beta $ 
phase transition was not found in our calculations, in accord 
with the experimental fact that the transitions are promoted by 
the disorder existing in the real samples \cite{meyer1}, if not, 
metastable states can be maintained for days. 
As calculated in refs. \onlinecite{claudio1,claudio2} with the isotropic 
model, also for the anisotropic atom-atom model the solid-liquid 
phase transition is located at temperatures near the 
experimental value ($\sim$400 K), but only for a disordered 
sample of S$_8$ molecules.
This sample is analized with a MD algorithm that allows 
changes in the MD box volume, but not in its angles, as a function 
of pressure and temperature. The disorder is generated by initially 
increasing the temperature of a cubic sample to 500 K and afterwards 
following its evolution by increasing and decreasing the temperature 
in the 350-450 K range. The result is totally similar to Fig. 4 of 
our ref. \onlinecite{claudio1} and it is not included here. 
 This last calculation is only of theoretical interest, and we did not
intend to simulate the real liquid phase of elemental sulfur.    
The experimental data show that the disorder, in the case of the 
real solid-liquid phase transitions, is generated because the S$_8$ 
molecules start to dissociate when the temperature of the sample is 
near the melting point and our calculations should be valid very near 
the transition, when the fraction of broken molecules is small.

\section{Conclusions} 

It has to be pointed out that, as we are dealing with molecular 
crystals, as long as no chemical reactions or electronic excitations 
are involved, the classical approximation is good enough for the 
calculation of structural and dynamical properties, provided that 
the inter- and intramolecular potential gives a good description 
of the real interactions. Moreover, simple model molecules can be 
very helpul to elucidate the complex phase diagram of the different
crystalline allotropes of elemental sulfur, some of them with 
a large number of molecules per cell and others with molecular 
reorientational disorder.\\

 In refs. \onlinecite{claudio1,claudio2} we verified that, surprisingly,
a large amount of crystalline properties can be very nearly
reproduced by a simple 
and isotropic atom - atom LJ model. The most discordant
point being the finding of a low temperature monoclinic sample
$\alpha $'-S$_8$, that has not been observed \cite{alfa2}
 but, nevertheless, might
 be a metastable phase of this elemental compound.\\

Assuming that this $\alpha $'-S$_8$ phase is, probably, an artifact 
of the simple isotropic intermolecular potential model, in this paper 
we analized two anisotropic case molecular models.\\

Our calculations show that an anisotropic ANI-S$_4$ model with local 
S$_4$ symmetry calculates the same stable structures and dynamical 
properties than the isotropic model, in the time scale of
our simulations (about 200 ps.), being the isotropic model more 
sensitive to changes in the thermodynamic parametes P and T than 
the anisotropic ANI-S$_4$ one.\\

The results obtained with the  ANI$_{\perp}$ case model are, instead,
 qualitatively very different from the other models. The four 
crystalline phases are calculated as polymorphic structures, but their
relative cohesive energy is changed. With this model molecule 
the $\alpha $-S$_8$ phase has a lower configurational energy  
than $\alpha $'-S$_8$ in all the studied range of temperatures,
in agreement with available measurements.
 The calculated density of $\gamma $-S$_8$ crystals is in accord
with X-ray measurements, but not with the density data 
of ref. \onlinecite{meyer1,gam1}.\\
 
  In conclusion, because of the amount of experimental data that these 
two case model
molecules nearly reproduce, and due to the fact that the $\alpha $'-S$_8$ 
allotrope 
is calculated stable (for about 200ps.) with three model potentials, 
we propose a thorough measurement of the crystalline
phases of S$_8$, including different techniques employed to grow the crystals  
at different temperatures. Additional measurements of the 
$\beta $- and $\gamma $-S$_8$ crystals are also needed, in particular
measurements of molecular reorientational times for $\beta $-S$_8$ 
and lattice
parameters as a function of temperature and pressure. Accurate 
measurements of the density of $\gamma $-S$_8$ are still lacking.\\

  Detailed calculations of the lone pair orientation in 
elemental sulfur molecules, including open chains and cyclic molecules
of different size, are presently being carried out using the
GAMESS (version 2001) \cite{gamess} program. Lastly, a further and 
simple anisotropic model will be developed with which we
 plan to extend our calculations to the crystalline phases of
the molecular allotropes and the liquid phase of elemental sulfur.\\
 
\begin{acknowledgements}
This work was partially supported by the grant PIP 0859/98 of CONICET and 
Fundaci\'on  J. Balseiro-2001. C.P. thanks partial support from FOSDIC.
\end{acknowledgements}

\newpage

\textbf{Table I:} Experimental and calculated lattice parameters: a, b, c (\AA)
and monoclinic angle $\beta $(deg.). $iso$ denotes the calculations performed 
with the isotropic atom-atom model\cite{claudio1,claudio2} 
and ANI those performed with the corresponding anisotropic
model. Z is the number of molecules 
in the cell, V (\AA $^3$) is the volume per molecule.
 See the text for a discussion about the $\gamma $-S$_8$ data.\\

\begin{tabular}{llllllll}
{\bf Phases}& &{\bf Z}&{\bf a(\AA )}&{\bf b(\AA )}&{\bf c(\AA )}&{\bf$\beta $(deg.)}&
{\bf V(\AA $^3$)}\\ 
&  &  &  &  &  &  &  \\ 
{\bf$\alpha $-S}$_8$ & Ref. \cite{alfa1} & 16 & 10.4646(1) & 12.8660(1) & 
24.4860(3) & - & 206.046(98) \\ 
(T=300 K) & $iso$ &  & 10.34(10) & 13.20(3) & 24.32(7) & - & 207.3(1.0) \\ 
(T=300 K) & ANI-S$_4$ ($d_{ani}$=0.1\AA) &  & 10.48(15) & 13.01(3) & 24.29(30) & - & 207.2(2) \\ 
(T=300 K) & ANI-S$_4$ ($d_{ani}$=0.6\AA) &  & 10.41(3) & 13.08(4) & 24.24(12) & - & 206.8(8) \\ 
(T=300 K) & ANI$_\perp$ ($d_{ani}$=0.6\AA) &  & 10.24(4) & 12.78(4) & 25.07(3) & - & 205.3(3) \\ 

{\bf$\beta $-S}$_8$ & Ref. \cite{meyer1} & 6 & 10.778 & 10.844 & 10.924 & 95.8$%
^{\circ }$ & 211.70 \\ 
(T=370 K) & $iso$ &  & 11.01(11) & 10.75(3) & 10.79(10) & 97.48(7)$ ^{\circ }$ & 211.3(7) \\ 
(T=375 K) & ANI-S$_4$ ($d_{ani}$=0.1\AA) &  & 11.00(1) & 10.80(6) & 10.80(9) & 96.89(4)$ ^{\circ }$ & 212.3(9) \\ 
(T=375 K) & ANI-S$_4$ ($d_{ani}$=0.6\AA) &  & 10.86(1) & 11.07(3) & 10.59(10) & 93.4(6)$ ^{\circ }$ & 211.0(3) \\ 
(T=375 K) & ANI$_\perp$ ($d_{ani}$=0.6\AA) &  & 10.61(2) & 10.92(2) & 10.94(2) & 92.7(1)$ ^{\circ }$ & 212.2(1) \\ 

{\bf$\gamma $-S}$_8$ & Ref. \cite{gam1} & 4 & 8.442(30) & 13.025(10) & 9.356(50)
& 124.98(30)$^{\circ }$ & 210.8(1.3) \\ 
(T=300 K) & $iso$ &  & 8.03(13) & 13.16(14) & 8.88(11) & 120(3)$^{\circ }$ & 202.5(2.5)\\
(T=300 K) & ANI-S$_4$ ($d_{ani}$=0.1\AA) &  & 8.03(10) & 13.16(4) & 8.89(8) & 121(2)$^{\circ }$ & 201.2(9) \\   
(T=300 K) & ANI-S$_4$ ($d_{ani}$=0.6\AA) &  & 8.04(9) & 13.30(13) & 8.96(8) & 121.3(8)$^{\circ }$ & 204.9(3) \\   
(T=300 K) & ANI$_\perp$ ($d_{ani}$=0.6\AA) &  & 8.21(4) & 13.05(5) & 9.11(5) & 122.9(4)$^{\circ }$ & 205.2(6) \\   

{\bf$\alpha $'-S}$_8$ &  & 16  &  &  &  &  &  \\   
(T=100 K) & $iso$ &  & 13.18(7) & 9.31(5) & 26.7(1) & 76.1(5) & 198.9(9) \\
(T=100 K) & ANI-S$_4$ ($d_{ani}$=0.1\AA) &  & 13.08(7) & 9.49(5) & 26.0(1) & 77.1(6) & 196.5(6) \\
(T=100 K) & ANI-S$_4$ ($d_{ani}$=0.6\AA) &  & 13.29(7) & 9.39(5) & 26.3(3) & 76.4(4) & 199.3(6) \\
(T=100 K) & ANI$_\perp$ ($d_{ani}$=0.6\AA) &  & 13.07(12) & 9.38(10) & 27.0(1) & 78.2(6) & 202.5(9) \\
\end{tabular}

\newpage

{\bf Figure:}

Fig. 1: Crystalline phases of the molecular S$_8$ compound. From left 
to right: a) orthorhombic $\alpha $-S$_8$ at 300 K, 
b) monoclinic  $\beta $-S$_8$ at 150 K (with the high temperature
orientationally disordered molecules in light colour), c) pseudo-hexagonal  
$\gamma $-S$_8$ at 300 K,
d) other view of the orthorhombic $\alpha $-S$_8$ at 300 K,
e) calculated monoclinic $\alpha $'-S$_8$, at 100 K, with the
isotropic LJ model \cite{claudio1,claudio2}.\\

Fig. 2: Configurational energy U$_{conf}$ (kJ/mol) and volume per 
molecule (\AA  $^3$) at 0 kbar and as a function of temperature T (K), 
as given by the anisotropic ANI-S$_4$ model, with
$d_{ani}$=0.1 \AA. The lines  
are a guide to the eyes.\\

Fig. 3: Configurational energy U$_{conf}$ (kJ/mol) and volume per
molecule (\AA  $^3$) at 0 kbar and as a function of temperature T (K),
as given by the anisotropic ANI$_\perp $ model, with
$d_{ani}$=0.6 \AA. The lines
are a guide to the eyes.  \\ \\ 

Fig. 4:  Gibbs free energy (kJ/mol) at 0 kbar 
and as a function of temperature T (K),
as given by the anisotropic ANI$_\perp $ model, with
$d_{ani}$=0.6 \AA. The lines
are a guide to the eyes.  \\ \\   
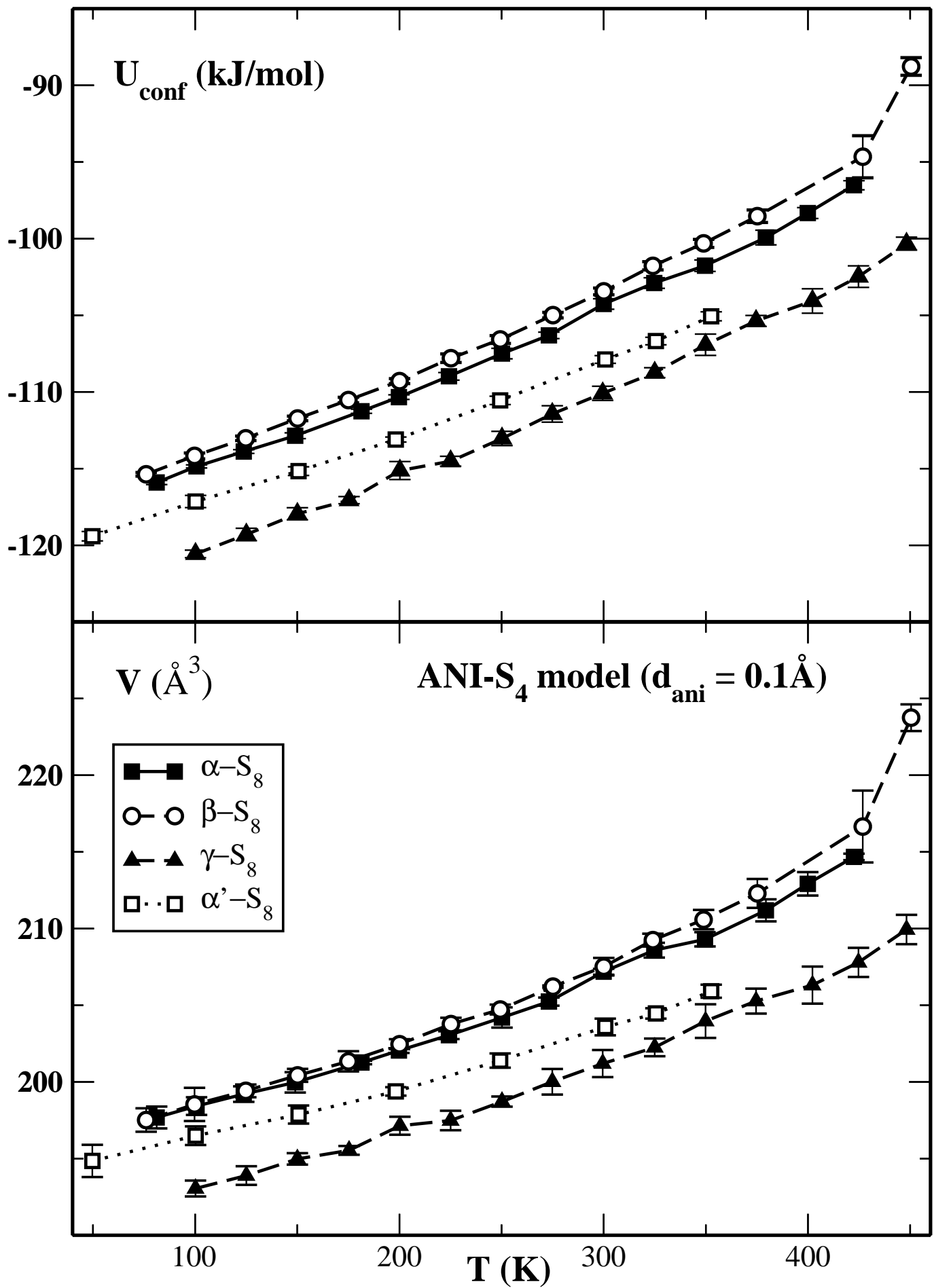
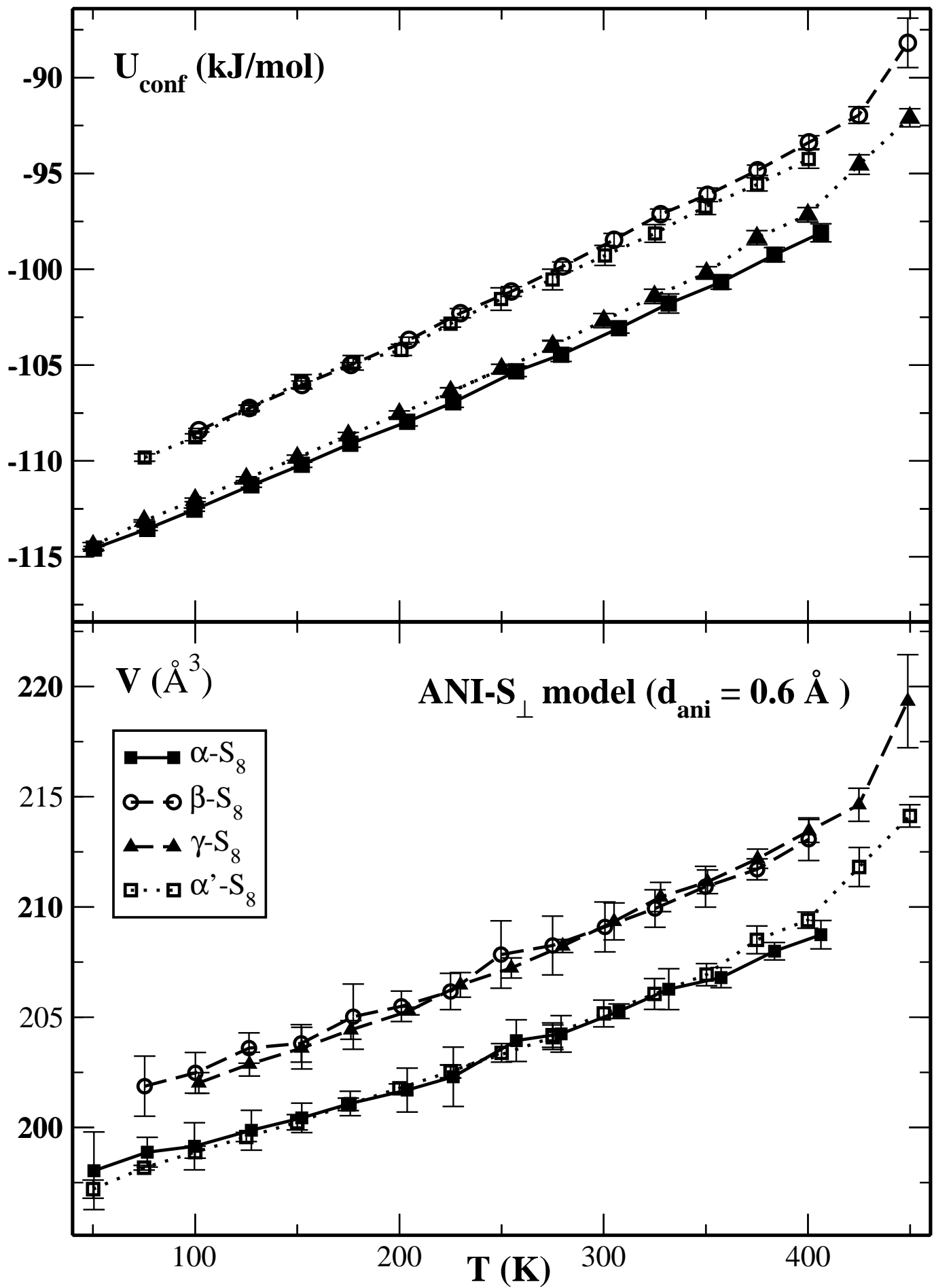
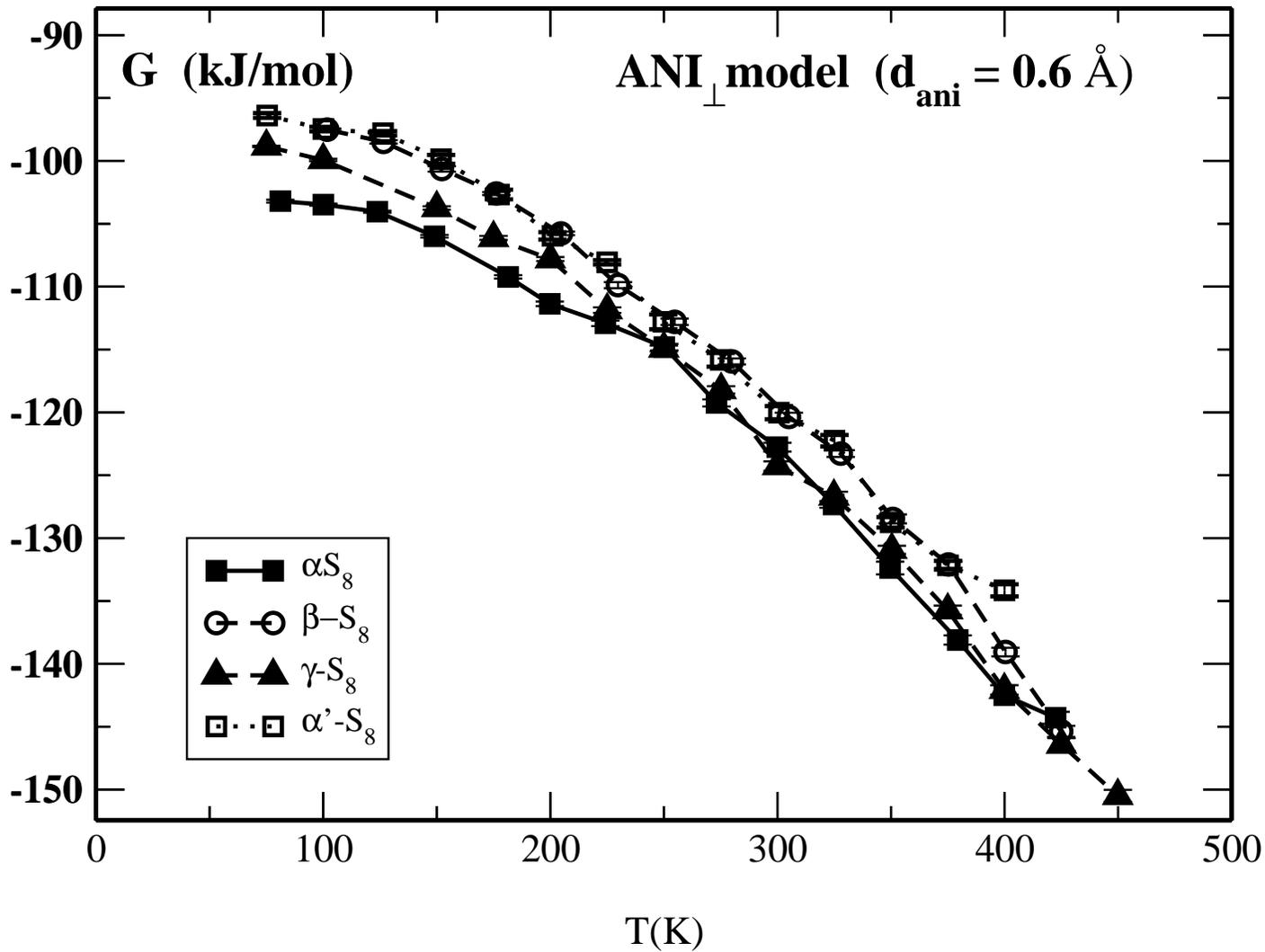
\end{document}